# Software Components for Web Services

[1]Muthu Ramachandran [2]T.R.Gopalakrsihnan Nair [3]R.Selvarani

[1]Innovation North: The faculty of Information and Technology
Leeds Metropolitan University, Leeds, LS6 3QS, UK
m.ramachandran@leedsmet.ac.uk
[2]Director- Research & Industry, trgnair@ieee.org
[3]Head Software Engineering Research - RIIC, CSE Department, selvss@yahoo.com
[2,3]Dayananda Sagar Institutions, Bangalore, India

**Abstract -** Service-oriented computing has emerged as the new area to address software as a service. This paper proposes a model for component based development for service-oriented systems and have created best practice guidelines on software component design.

## I. Introduction

Service oriented software engineering has emerged to address software as a set of services which deal with user needs. Therefore a service may make call requests to a range of systems (integrated software applications) and hence the emphasis is on design flexibility. We need to understand the concept behind services and components before making our design choice. The main emphasis for choosing various design rationale is to manage change and flexibility in software design and to meet user requests for a range of random services. The issues of managing change support two main perspectives, business and technical. Flattery (2007) says software must constantly evolve to meet constantly changing business requirements.

There are many ways to define and implement services. Traditionally a service is invoked by direct human users which may not be the best definition with respect to current service-oriented paradigm which has emerged in recent years. In practice, services may provide capability for immediate use based on well-defined behaviour, inputs, and outputs. Services should be managed to meet quality and other non-functional objectives. Services are also organised to meet overall organisational goals such as business, customer satisfaction, social, employee training needs, and financial. Characteristics of services are modelable, and composable (services made up of services) and these are similar to characteristics of open distributed and telecom frameworks. Service oriented computing allows to move from unmeasured to measured economy by means of measuring service invocation, input versus output (GDP), employer sector, and employee activities.

Service oriented computing involves several relationships amongst at least three major subject areas such as *human activity* (including study of psychology, knowledge sharing, requirements engineering, social sciences, understanding people and culture, ethnography), organizational and business processes (detailed study of management science, organisational structures, external and customer relationship management system), and web and computational services (detailed analysis of web enabled services, computational aspects, design attributes such as run-time re-configurability, integration, testing, componentisation).

It is worth considering some examples of types of service-oriented applications such as e-health, e-energy, e-government, e-business, e-commerce, e-learning, e-web services, and e-science. The following section provides a framework for evaluation of component-based versus service-based paradigm. There is a complete paradigm shift in design and development. Elfatatry (2007) says software must constantly evolve to meet constantly changing business requirements.

The notion of a *publish/subscribe paradigm* proposed is a basis for middleware platforms that support software applications composed of highly evolvable and dynamic federations of components. This paradigm decouples the communication among components and supports implicit bindings among components.

Grid computing provides facilities to enable sharing, selection, and aggregation of resources distributed across multiple administrative domains based on the resources' availability, capacity, performance, cost, and QoS (Quality of service) requirements.

Autonomic computing is another emerging area which provides features for self-configuration, self-healing, self-optimization and self-protection to manage with complexity, heterogeneity, and uncertainty of modern software systems. Researchers are also studying new interaction and cooperation paradigms inspired by biological systems. In particular, they are modelling the abilities of single biological entities (for example, ants) to self-organize and behave autonomously and together achieve a common objective (Baresi et al. 2006). In essence (Service Oriented Architechture) *SOA = semantic integration + Loose coupling + Managed evolution;* Semantic integration is the major prerequisite, loose coupling is the distinct feature of SOA, and managed evolution represents both a purpose and an implementation approach (Helbig 2007).

Service-oriented software systems build on the SOA paradigm which proposes to decouple service requirements from service implementation in terms of platform, location, availability, and versions. SOA concentrates on specification and implementation of services as entities and mainly focuses on web services. Systems composed of services represent a major shift from standard software components and services can support open-world and distributed software development at a reduced cost (Nano and Zisman 2007). To tackle all such challenges we need a much more promising approach to software architectures such as service-oriented architectures supporting these types of applications, one of which is proposed by Baresi et al. (2006).

There have been several interesting research work published in this area. For example, Baresi et al (2006) have identified a list of supporting technologies which might be a useful source for SoC developers. They have also provided a good set of resources on methods, techniques, and technologies required for implementing services. We need a systematic approach and process to develop services, which is described in the following section.

**Characteristics of service-oriented systems**

Identifying characteristics of a services-oriented system is vital for designers so that they can select, design, and evaluate those characteristics that are applicable to their applications. In SoA some of the characteristics will overlap since it involves integration of several disciplines and subject areas.

Some of the identified services and component characteristics are:

- Reusable web services and other core services
- Enterprise integration services
- Dynamic binding and reconfigurable at run-time
- Granularity
- Publish, Subscribe, and Discover
- Open-world where components must be able to connect and pluggable to third party software systems or components.
- Heterogeneity supporting cross-platform applications
- Re-configurable
- Self-composable and recoverable
- Grid infrastructure and resources management
- Autonomic framework
- Middleware
- QoS

This is illustrated in Figure 1 which shows some of the above characteristics that are the key components for software development. In modern software development, characteristics such as open-world the components are customisable and connectable to third party systems. These components, and heterogeneity are crucial to developing highly reusable web services that will apply across domains and services.

The main reason for presenting such characteristics is to understand the basis for service-oriented systems and provide good practice design guidelines. The next section looks at the distinct features and differences between services and components. Again these characteristics need to overlap as we are also interested in applying component based development for service-oriented systems. In particular web services need to possess both services and components characteristics.

**Service Oriented Software Development Process**

Identifying service requirements requires a new (Reverse Engineering) RE process and modeling techniques as it is highly dependant on multi-level enterprises across corporations and enterprises. Identifying and knowing the complete requirements for all expected and unexpected services is very hard. The idea in service oriented engineering is to automatically publish automatically new services whereby service agents can then be able request and take advantage of required services for their customers. Figure 2 shows a development process model for service-oriented computing where initial requirements are captured based on enterprise-wide techniques and using domain analysis which should focus on a family of products and services. The second phase (Services RE) involves identifying a set of requirements of system services. This process involves service modelling and service specification for which we can use any well known techniques such as use case design and a template for service level specifications.

The third phase (Categorizing services) involves classifying and distinguishing services into various categories such as enterprise integration services (services across corporations, departments, other business services), software services which represent core functionality of software systems, business logic services which represents business rules and its constraints, web services (a self-contained and web enabled entity which provide services across customisable businesses at run-time), IT core services which include resource management, help desk systems, IT infrastructure, procurement, and delivery services, B2B and B2C services, Data services, QoS services, middleware services, Transaction management services, process integration services, re-configurability services, and grid services which include grid resource management and re-configurations.

**Proposed solution and software components**

The common pattern is to access excel and other documents and serving to user requests across the enterprise. The excel based software + service architecture is shown in Figure 3. This consists of a calculation definition request submitted to the calculation engine platform which consists of various core services (security, validation, scheduling, auditing, notification, and reporting). This subsequently publishes the results on the management dashboards and also produces automatically generated written documents.

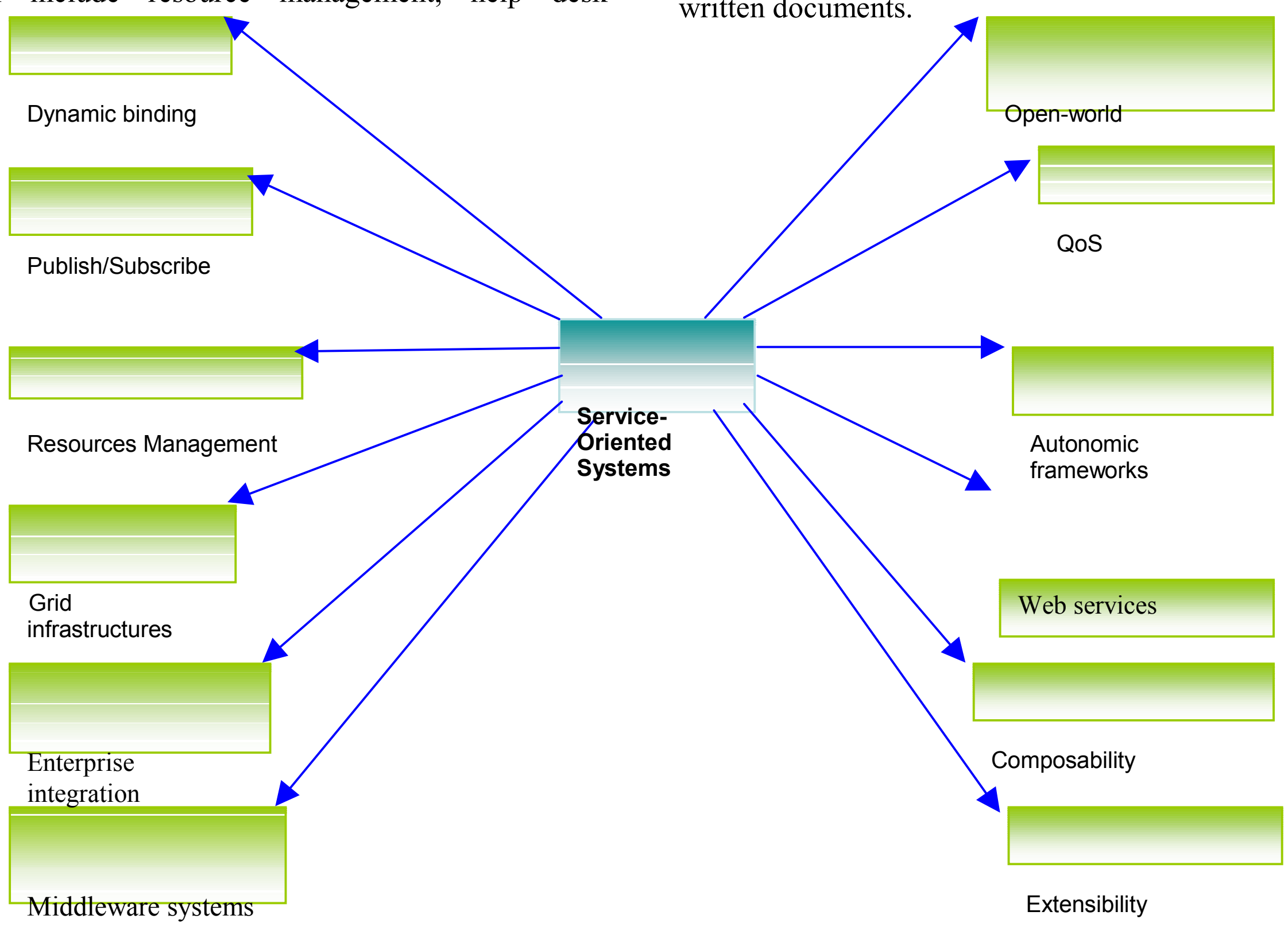

**Figure 1** Characteristics of service-oriented systems

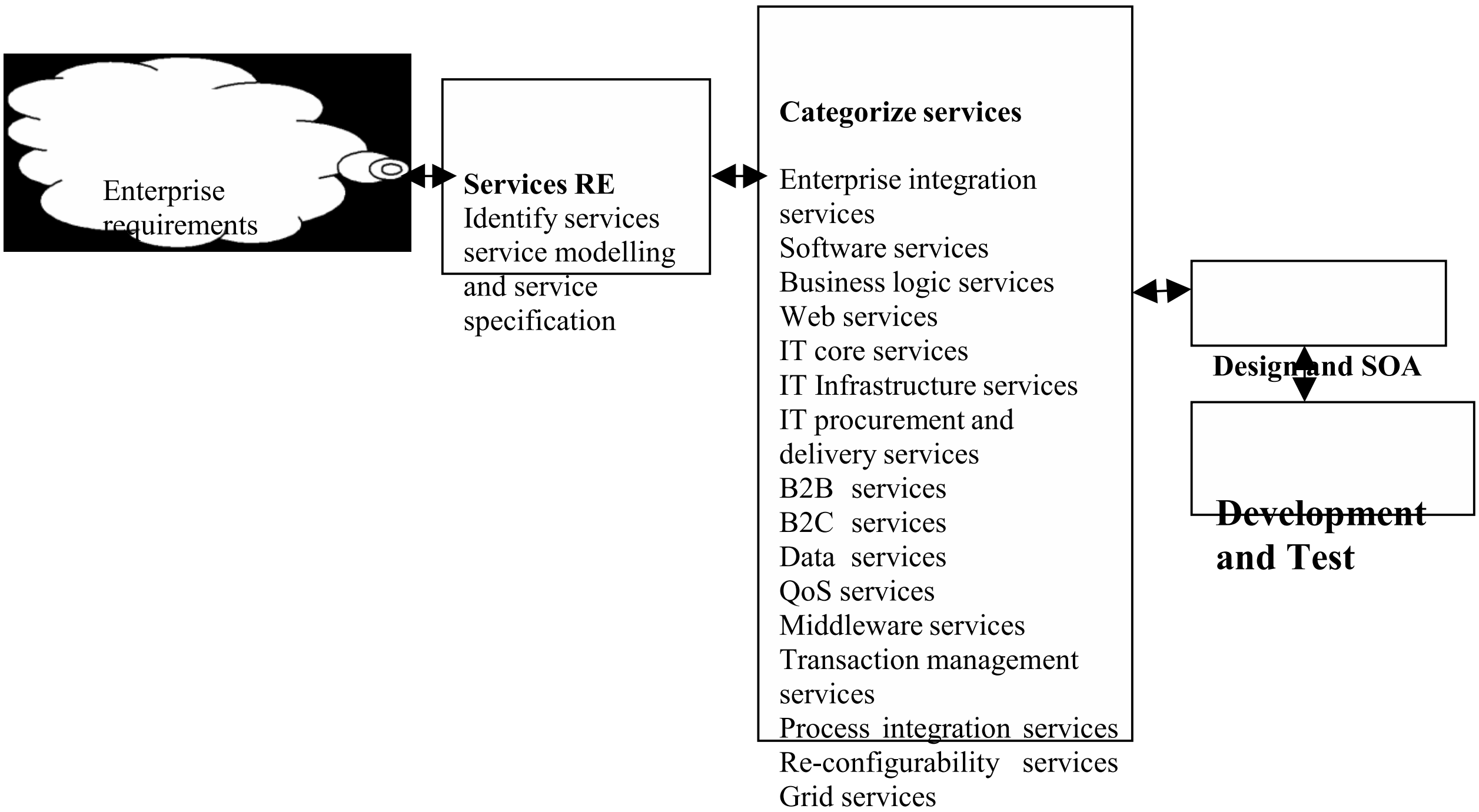

**Figure 2** Service-oriented software development process

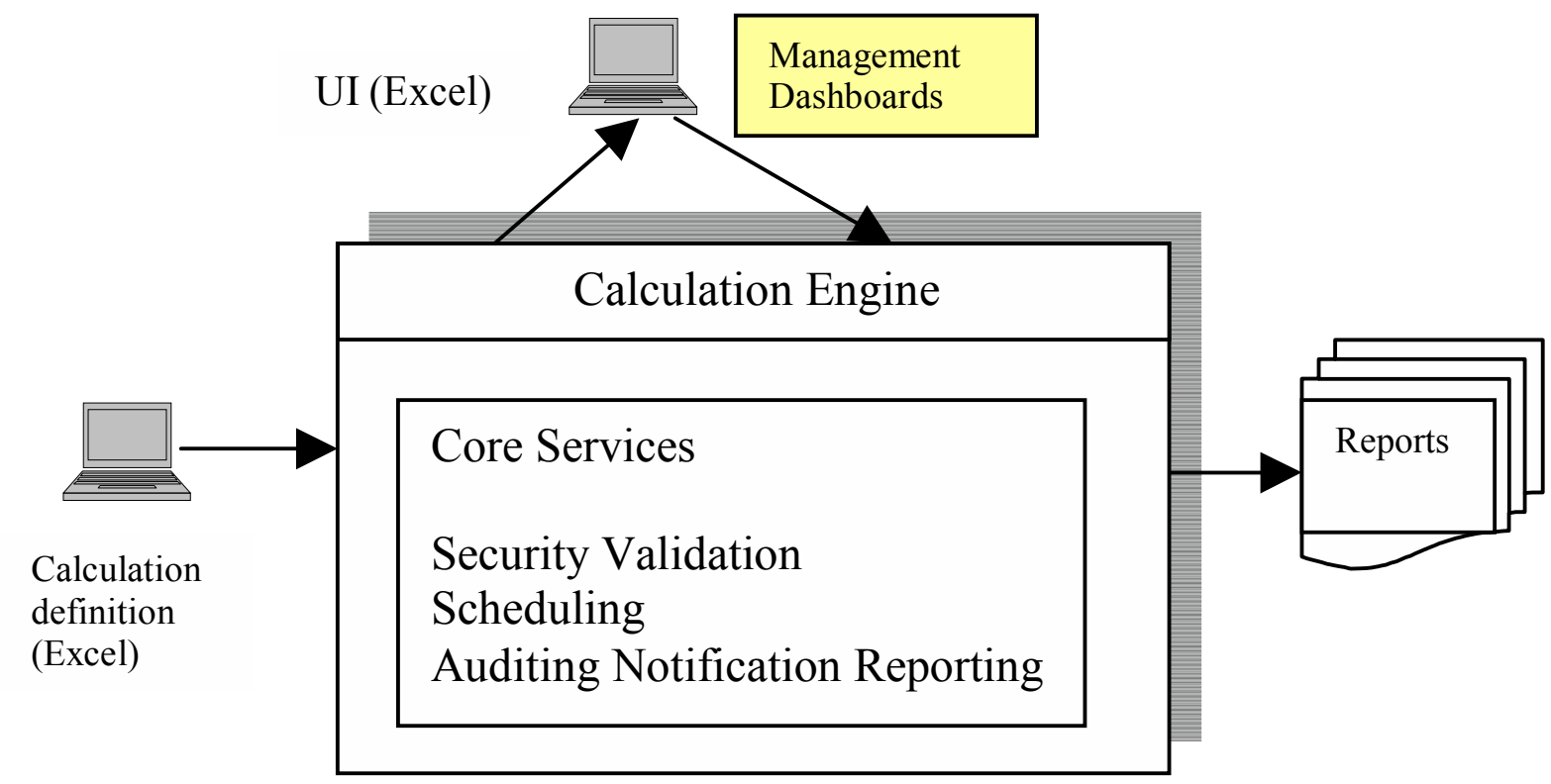

**Figure 3** Excel enterprise architecture for software + services

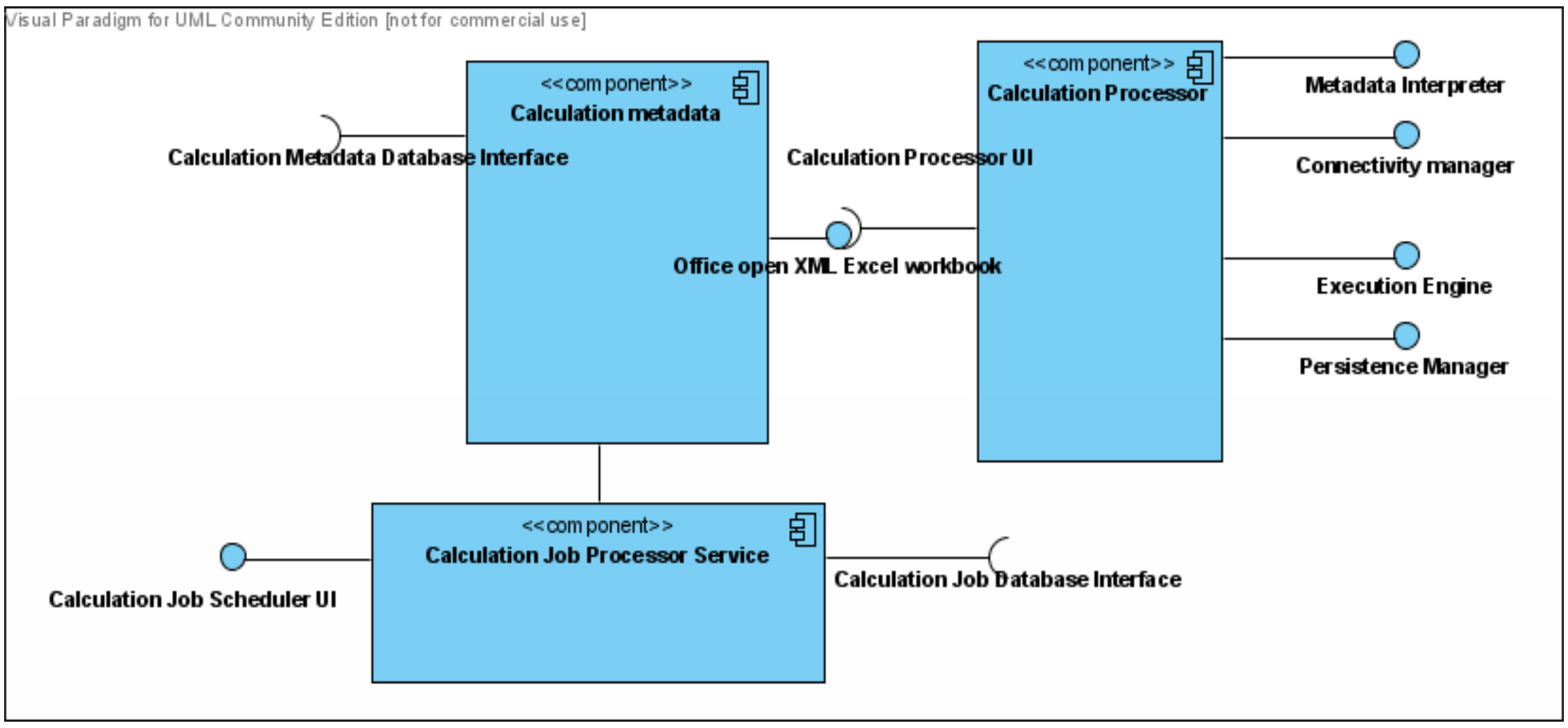

**Figure 4** UML components for business logic framework

This architectural pattern is quite similar to Model-View-Controller pattern (MVC) which has been a popular pattern in user interface design and other applications. Generalising these three solutions yields the software + services architectural pattern, an OBA solution utilising Excel as a client connected to distributed enterprise systems using web services. Web services components that are shown in Figure 4 retrieve current information immediately into the user's worksheet. Authorised users can make changes to (Electronic Services) ESs which will be available across the entire enterprise. All charts, graphs, tables, and other reporting tools that reference the changed data are automatically updated to reflect the current and precise information. Figure 4 provides a component based solution for business logic framework. This consists of three components - Calculation metadata, Calculation processor, and Calculation job processor services supporting both provider and require interfaces.

A versatile business logic framework allows workbook designers (skilled Excel users who understand a workbook's data relationship and their presentation) to change the system's business rules without bringing in a developer to recompile and redeploy the code. The list of components and interfaces shown provides the following functionalities and performs certain tasks:

- A Calculation processor for configuring, saving, and executing a calculation job
- A calculation processor UI implements calculation processor component's user interface
- A calculation job processor service for running calculations on the server using Excel services invoked from a Windows service
- A calculation job scheduler for configuring the time-based execution of calculation jobs

These are some of the rationale for the technical design solution which has been provided and can be extended and reused. Service oriented software engineering is a growing area of research and there are several challenges ahead. The above lists have identified a few of the current research issues in SoC that dominate the field today. This is a multi-disciplinary field and hence software engineers need to work closely with business experts and also need to revise curriculum development for the future accordingly.

## Conclusion

Service oriented computing has emerged and will dominate a new era in computer science.. Software components will play a major role in this area since it requires web services components for multi-connectivity and re-configurability. All of which can be achieved by using careful design of software components.